\documentstyle[12pt,epsf]{article}

\begin{document}

\begin{titlepage} 
\begin{center} 
 
{\Large \bf Boundary dependence of the coupling constant and the mass in the
vector $N$-component $(\lambda\varphi^{4})_{D}$ theory}\\
\vspace{.3in} 
{\large\em A.P.C. Malbouisson${}^{(a)}$, J.M.C. Malbouisson${}^{(b)}$} \\
\vspace{0.1cm} 
${}^{(a)}$ {\it Centro Brasileiro de Pesquisas F\'{\i}sicas,} \\ 
\vspace{0.1cm} 
{\it Rua Dr. Xavier Sigaud 150, Urca,} \\ 
\vspace{0.1cm} 
{\it Rio de Janeiro CEP 22290-180-RJ, Brazil.} \\ 
\vspace{0.1cm}  
{\it E-mail: adolfo@lafex.cbpf.br} \\ 
\vspace{0.7cm} 
${}^{(b)}$ {\it Instituto de F\'{\i}sica, Universidade Federal da Bahia,} \\ 
\vspace{0.1cm} 
{\it Campus Universitario de Ondina, 40210-340-BA Salvador Brazil} \\ 
\vspace{0.1cm}  
{\it E-mail: jmalboui@ufba.br} \\ 
\vspace{0.7cm}

\end{center} 
 
\subsection*{Abstract} 
Using the Matsubara formalism,  
we consider the massive $(\lambda\varphi^{4})_{D}$ vector $N$-component model
in the large $N$ limit, the system being confined between two infinite
paralell planes.  We investigate the behaviour of the coupling constant as 
a function of the separation $L$ between the planes. For the Wick-ordered
model in $D=3$ we are able to give an exact formula to the $L$-dependence 
of the coupling constant. For the non-Wick-ordered model we indicate how
expressions for the coupling constant and the mass can be obtained 
for arbitrary dimension $D$ in 
the small-$L$ regime. Closed exact formulas for the $L$-dependent 
renormalized coupling constant 
and mass are obtained in $D=3$ and their behaviours 
as  functions of $L$ are displayed. We are also able to obtain in generic
dimension $D$, an equation for the critical value of $L$ corresponding to
a second order phase transition in terms of the Riemann  $zeta$-function.
In $D=3$ a renormalization is done and an explicit formula for the critical
$L$ is given.

\end{titlepage} 
\newpage \baselineskip.37in

\section{Introduction}

Translationally non-invariant field theories are of great interest in
several branches of theoretical physics, particularly in the study of
confined systems, or systems undergoing changes of state induced by space
inhomogeneities. For instance, in the Ginzburg-Landau theory of phase
transitions applied to the study of type II superconductors, the
external magnetic field, constant in some fixed spacial direction breaks
translational invariance introducing important changes in the Feynman rules 
\cite{Lawrie1}, \cite{Lawrie2}, \cite{Brezin}. Another important situation
is found in the finite temperature field theoretical formalism of
statistical mechanics, in particular the Matsubara imaginary time formalism,
which introduces a breaking of translational invariance along the imaginary
time (temperature) axis, leading to corresponding modifications in the
Feynman rules. In this context, in recent papers (\cite{NAdolfo1}, \cite
{NAdolfo2}, \cite{Gino1}) the behaviour of coupling constants on the
temperature has been investigated in connection to stability and phase
transitions. In \cite{NAdolfo2} the finite temperature behaviour of the $ 
\lambda \varphi _{D}^{4}$ model was investigated at one-loop approximation.
One of the results is that the thermal squared mass (in the disordered phase)
is a positive increasing function of the temperature, while the thermal
correction to the renormalized coupling constant is negative and increasing
in modulus with the temperature. This could be interpreted as the possibility
of the vanishing
of the thermal coupling constant at some temperature and its change of sign
afterwards. Of course we can not conclude from this one-loop result that there
is a breaking of stability at some temperature, since we can not 
be sure that this peculiar behaviour will
be preserved when higher order loop's contributions are considered (see
below).

Questions concerning stability and the existence of phase transitions may
also be raised if we consider the behaviour of field theories as function of
spacial boundaries. In this situation, for Euclidean field theories, time
(temperature) and the spacial coordinates are on exactly the same footing
and the Matsubara formalism applies as well for the breaking of invariance
along any one of the spacial directions. The possibility of vanishing of the
coupling constant as mentioned above for temperature, or more generally the
existence of phase transitions, would be in this case associated to values
of some spacial parameters describing the breaking of translational
invariance, for instance the distance $L$ between planes confining the
system. In this case, in a very similar way as it has been remarked in the
context of finite temperature field theory in \cite{Gino1}, two-loop
corrections in the small $L$ regime are positive and tend to compensate the
one-loop lowering of the value of the $L$-dependent renormalized coupling
constant. Similar questions could be raised in general about the existence
of phase transitions induced by spacial boundaries separating different
states of the system. One of the most striking examples of this kind of
situation is the spacial separation between the Abrikosov lattice of
vortices and the other phases in type II superconductors \cite{Abrikosov}.
In another related domain of investigation, there are systems that present
defects, created for instance in the process of crystal growth by some
prepared circumstances (domain walls). At the level of effective field
theories, in many cases this can be modelized by considering a Dirac
fermionic field whose mass changes of sign as it crosses the defect, what
means that the domain wall plays the role of a critical boundary, separating
two different states of the system 
(see for instance \cite{Cesar}, \cite{FAdolfo} and other references
therein). In any case
the question of the behaviour of physical parameters in field theories
(masses or coupling constants) on spacial boundaries or inhomogeneities, is
still largely open and deserves further investigation.

In order to shed some
light on the problem, in this article we examine a model that allows a
non-perturbative approach. We consider the Euclidean vector $N$-component $ 
\lambda\varphi^{4}_{D}$ theory at leading order in $\frac{1}{N}$, the system 
being confined between two infinite parallel planes a distance $L$ 
apart from one another. In this
case we are able to obtain for the Wick-ordered model an exact formula for
the renormalized coupling constant behaviour as a function of the given
spacial boundaries. For the non-Wick-ordered model we obtain for the
coupling constant and the mass, formulas valid in the small-$L$ regime for
arbitrary dimension $D$. In dimension $D=3$ we obtain 
 closed formulas for the renormalized $L$-dependent coupling 
constant and mass and plots of these parameters as 
functions of $L$ are given. We are also able to obtain in generic dimension
$D$, an equation for
the critical value of $L$ corresponding to a second order phase transition
in terms of the Riemann $zeta$-function. In $D=3$ a renormalization is done
and an explicit formula for the critical $L$ is given.

\section{The Wick-ordered model}

We consider the model described by the Lagrangian density,
\begin{equation}
{\mathcal L}=\frac{1}{2}\partial_{\mu}\varphi_{a}\partial^{\mu}\varphi_{a}
+\frac{1}{2}m^{2}:\varphi_{a}\varphi_{a}:+\frac{\lambda_{0}}{N}
(:\varphi_{a}\varphi_{a}:)^{2},
\label{Lagrangeana}
\end{equation}
in Euclidean $D$-dimensional space, where $\lambda _{0}$ is the bare
coupling constant, $m$ is the physical mass (see below) and summation over
repeated "colour" indices $a$ is assumed.  We note that Wick-ordering of
products of fields makes
unnecessary to perform an explicit mass renormalization at the order $\frac{ 
1}{N}$ at the one-loop approximation. The tadpoles are suppressed by
Wick-ordering. Thus we consider in the following only the zero external
momenta four-point function, which we take as the basic object for 
our definition of the
renormalized coupling constant. To simplify the notation in the following we
drop out the colour indices, summation over them being understood in 
field products.

The four-point function at leading order in $\frac{1}{N}$ is given by the
sum of all the diagrams of the type depicted in $Fig.1$. We consider the system 
confined between two
parallel planes a distance $L$ apart from one another, normal to the $x$
-axis. We use Cartesian coordinates ${\bf r}=(x,{\bf z})$, where ${\bf z}$
is a $(D-1)$-dimensional vector, with corresponding momenta ${\bf k}=(k_{x}, 
{\bf q})$, ${\bf q}$ being a $(D-1)$-dimensional vector in momenta space.
Then since the $x$-dependence of the field is confined to a segment of
length $L$, we are allowed to perform the Matsubara replacements, 
\begin{equation}
\int \frac{dk_{x}}{2\pi }\rightarrow \frac{1}{L}\sum_{n=-\infty }^{+\infty
}\;,\;\;\;\;\;\;k_{x}\rightarrow \frac{2n\pi }{L}\equiv \omega _{n}.
\end{equation}
Performing the sum over all diagrams indicated in $Fig.1$ we get for the $L$
-dependent renormalized four-point function at zero external momenta, 
\begin{equation}
\Gamma _{D}^{(4)}(0,L)=\frac{1}{N}\;\;\frac{\lambda _{0}}{1-\lambda
_{0}\Sigma (D,L)},  \label{4-point1}
\end{equation}
where $\Sigma (D,L)$ corresponds to the single bubble diagram in $Fig.1$,
after subtraction of the ultraviolet divergent parts.

To obtain an expression for $\Sigma (D,L)$, we briefly recall some one-loop
results described in \cite{NAdolfo2} adapted to our present situation. These
results have been obtained by the concurrent use of dimensional and
$zeta$-function
analytic regularizations. In order to deal with dimensionless
quantities in the regularization procedures, we introduce parameters $ 
c^{2}=m^{2}/4\pi ^{2}\mu ^{2},\;\;\;\;(L\mu )^{2}=a^{-1},\;\;\;g=(\lambda
/8\pi ^{2}),\;\;\;(\varphi _{0}/\mu )=\phi _{0}$, where $\varphi _{0}$ is
the normalized vacuum expectation value of the field (the classical field)
and $\mu $ is a mass scale. In terms of these parameters the
boundary-dependent ($L$-dependent) one-loop contribution to the effective
potential can be written in the form, 
\begin{equation}
U_{1}(\phi ,L)=\mu ^{D}\sqrt{a}\sum_{s=1}^{\infty }\frac{(-1)^{s}}{2s}
g^{s}\phi_{0} ^{2s}\sum_{n=-\infty }^{+\infty }\int \frac{d^{D-1}k}{(an^{2}
+c^{2}+{\bf k}^{2})^{s}},  \label{potefet1}
\end{equation}
or, using a well-known dimensional regularization formula \cite{Zinn}, 
\begin{equation}
U_{1}(\phi ,L)=\mu ^{D}\sqrt{a}\sum_{s=1}^{\infty }f(D,s)g^{s}\phi_{0}
^{2s}Z_{1}^{c^{2}}(s-\frac{D-1}{2};a),  \label{potefet2}
\end{equation}
where $f(D,s)$ is a function proportional to $\Gamma (s-\frac{D-1}{2})$ and $ 
Z_{1}^{c^{2}}(s-\frac{D-1}{2};a)$ is one of the Epstein-Hurwitz $zeta$
-functions defined by, 
\begin{equation}
Z_{K}^{c^{2}}(u;a_{1},...,a_{K})=\sum_{n_{1},...,n_{K}=-\infty }^{+\infty
}(a_{1}n_{1}^{2}+...+a_{K}n_{K}^{2}+c^{2})^{-u},  \label{zeta}
\end{equation}
valid for $Re(u)>\;K/2$ (in our case $Re(s)>\;D/2$). The Epstein-Hurwitz $ 
zeta$-function can be extended to the whole complex $s$-plane and we obtain,
after some manipulations \cite{Elizalde}, the one-loop correction to the
effective potential, 
\begin{equation}
U_{1}(D,L)=\frac{1}{4}\sum_{s=1}^{\infty }g^{s}\phi_{0} ^{2s}h(D,s)\left[
\Gamma
(s-\frac{D}{2})(\frac{m^{2}}{2\mu ^{2}})^{\frac{D-2s}{2}}+\sum_{n=1}^{\infty
}(\frac{m}{\mu ^{2}nL})^{\frac{D}{2}-s}K_{\frac{D}{2}-s}(mnL)\right] ,
\label{potefet3}
\end{equation}
where $h(d,S)=\frac{1}{2^{D/2-s-1}\pi ^{2s-D/2}}\frac{(-1)^{s+1}}{\Gamma
(s+1)}$ and $K_{\nu }$ are the Bessel functions of the third kind.

Coming back to the situation under study, we see that the single bubble
function $\Sigma (D,L)$ is the coefficient of the fourth power of the field
($s=2$) in the above equation. We can write $\Sigma (D,L)$ in the form, 
\begin{equation}
\Sigma (D,L)=H(D)-G(D,L),  \label{Sigma}
\end{equation}
where 
\begin{equation}
H(D)=\frac{1}{8\pi ^{2}}\frac{\pi ^{2s-D/2}m^{D-4}(-1)^{s+1}}{
2^{D/2-3}\Gamma (s+1)}\Gamma (2-\frac{D}{2})
\end{equation}
and 
\begin{equation}
G(D,L)=\frac{3}{2}\frac{1}{(2\pi )^{D/2}}\sum_{n=1}^{\infty }\left[ \frac{m}{
nL}\right] ^{D/2-2}K_{\frac{D}{2}-2}(mnL).  \label{G}
\end{equation}
>From the properties of Bessel functions it can be seen from Eq.(\ref{G})
that for any dimension $D$ $G(D,L)$ satisfies the conditions, 
\begin{equation}
\lim_{L\rightarrow \infty }G(D,L)=0\;\;\;\;\;\;\;\lim_{L\rightarrow
0}G(D,L)\rightarrow \infty .  \label{G1}
\end{equation}
We also conclude from Bessel functions properties that $G(D,L)$ is always
positive for any values of $D$ and $L$.

Let us define the renormalized coupling constant $\lambda $ in the absence
of boundaries as, 
\begin{equation}
\frac{1}{N}\lambda =\lim_{L\rightarrow \infty }\Gamma _{D}^{(4)}(0,L).
\label{lambdafree}
\end{equation}
>From Eqs.(\ref{4-point1}) and (\ref{Sigma}) we get, 
\begin{equation}
\Gamma _{D}^{(4)}(0,L)=\frac{1}{N}\;\;\frac{\lambda _{0}}{1-\lambda _{0}  
\left[ H(D)-G(D,L)\right] }  \label{4-point2}
\end{equation}
and from (\ref{lambdafree}), (\ref{G1}) and (\ref{4-point2}), 
\begin{equation}
\lambda =\frac{\lambda _{0}}{1-\lambda _{0}H(D)}.  \label{lambdafree1}
\end{equation}
Combining (\ref{lambdafree1}) and (\ref{4-point2}), we obtain the $L$
-dependent renormalized coupling constant, 
\begin{equation}
\Gamma _{D}^{(4)}(0,L)\equiv \frac{1}{N}\lambda (D,L)=\frac{1}{N}\;\frac{
\lambda }{1+\lambda G(D,L)}.  \label{lambda}
\end{equation}
The above procedure corresponds on perturbative grounds, to sum up all the
chains of bubble graphs in $Fig.1$ with all possible combinations of $\frac{
\lambda }{N}$ and counterterms $\frac{1}{N}\delta \lambda =\frac{\lambda
_{0}-\lambda }{N}$ at the vertices. It is just the resummation of all the
perturbative contributions including the counterterms from the chains of
bubbles and the subtraction of the divergent (polar) parts, written in
compact form. These subtractions are to be performed even in the case of odd
dimension, where no poles of $\Gamma $-functions are present.

An exact result can be obtained in dimension $D=3$. From the relationships 
\cite{Abramowitz}, 
\begin{equation}
K_{n+\frac{1}{2}}(z)=K_{-n-\frac{1}{2}}(z),\;\;\;\;\;K_{\frac{1}{2}}(z)= 
\sqrt{\frac{\pi }{2z}}e^{-z},  \label{Abramov}
\end{equation}
we obtain from Eqs.(\ref{lambda}) and (\ref{G}), after summing a geometric
series, the exact expression for the coupling constant in the large $N$
limit, 
\begin{equation}
\lambda _{W}(D=3,L)=\frac{8\pi m\lambda (e^{mL}-1)}{8m\pi
(e^{mL}-1)+3\lambda },  \label{lambdaexata}
\end{equation}
where the subscript $W$ is used to indicate explicitly Wick-ordering. A plot
of $\lambda _{W}(D=3,L)$ is given in $Fig.2$.

\section{The Non-Wick-ordered model}

The effect of suppression of Wick-ordering is that the renormalized mass can
not be taken as the coefficient of the term $\varphi_{a}\varphi_{a}$ in the
Lagrangian, as we have done in the preceding section. We must take a $L$
-corrected renormalized mass which satisfies the gap equation in the ordered
phase, 
\begin{equation}
m_{R}^{2}(L)=m^{2}+ \frac{4\lambda(N+2)}{N(2\pi)^{D/2}}\sum_{n=1}^{\infty}  
\left[\frac{m_{R}(L)}{Ln}\right]^{D/2-1}K_{\frac{D}{2}-1}(m_{R}(L)nL),
\label{mDyson}
\end{equation}
where $m$ stands for the constant mass parameter in the Lagrangian density
(\ref{Lagrangeana}), which corresponds to the renormalized mass in the
absence of
boundaries (free space) for the Wick-ordered model. To obtain the $L$
-dependent coupling constant the constant mass parameter $m$ should be
replaced in
Eqs(\ref{lambda}) and (\ref{G}) by the $L$-corrected mass $m_{R}(L)$ and the
resulting system of equations should be solved with respect to $L$. Under
these conditions  exact closed expressions are almost impossible to obtain,
since it would require a procedure equivalent to solving exactly the
Dyson-Schwinger equations.

Nevertheless results can be obtained, both analytic and numerical. In this
sense in the next we indicate how to generalize to our case in arbitrary
space dimension $D$, some results already obtained in refs. \cite{jac} and
\cite{tet} for $D=4$ in the context of thermal field theory. We take an
integral representation for the Bessel function in Eq. (\ref{mDyson})
\cite{Abramowitz}, 
\begin{equation}
K_{\nu}(z)=\frac{\sqrt \pi}{\Gamma(\nu +\frac{1}{2})}(\frac{z}{2}
)^{\nu}\int_{1}^{\infty}e^{-zt}(t^{2}-1)^{\nu-\frac{1}{2}}dt,
\label{repr.Bessel}
\end{equation}
valid for $Re(\nu)>-\frac{1}{2}$ and $|arg(z)|<\frac{\pi}{2}$. Using
this representation in Eq.(\ref{mDyson}), after some straightforward
calculations the $L$-dependent renormalized mass can be written in the
form, 
\begin{equation}
m^{2}_{R}(L)=m^{2}+\frac{4\lambda (N+2)}{N}F(D)m_{R}^{D-2}(L)
\int_{m_{R}(L)L}^{\infty} \frac{d\tau}{m_{R}(L)L}\left((\frac{\tau}{ 
m_{R}(L)L })^{2}-1\right)^{\frac{D-3}{2}} \frac{1}{e^{\tau}-1},
\label{mDyson1}
\end{equation}
where $F(D)=\frac{1}{2^{D}}\frac{1}{\pi^{\frac{D-1}{2}}} \frac{1}{\Gamma( 
\frac{D-1}{2})}$. When $D$ is odd, the power $\frac{D-3}{2}$ is an integer
and Newton binomial theorem gives a direct way for
evaluating the integral in Eq.(\ref{mDyson1}), implying in an algebraic
equation for
$m_{R}^{2}(L)$. When $D$ is even the expansion of $\left((\frac{
\tau}{m_{R}(L)L})^{2}-1\right)^{\frac{D-3}{2}}$ yields an infinite power
series. We obtain,

\begin{equation}
m_{R}^{2}(L)=m^{2}+\frac{4\lambda (N+2)}{NL^{D-2}}\sum_{k=0}^{\infty
}g(D,k)(m_{R}(L)L)^{2k}\int_{m_{R}(L)L}^{\infty }d\tau \frac{\tau ^{D-3-2k}}{
e^{\tau }-1}  \label{mDyson2}
\end{equation}
where $g(D,k)=F(D)(-1)^{k}C_{\frac{D-3}{2}}^{k}$ and the C's are the
generalized binomial coefficients for arbitrary power. For small values of $ 
k $ the integral that appears in the above equation is the well known Debye
integral 
\begin{equation}
I(x,n)=\int_{x}^{\infty }d\tau \frac{\tau ^{n}}{e^{\tau }-1}
=\sum_{q=1}^{\infty }e^{-qx}\;x^{n}\left( \frac{1}{q}+\frac{n}{q^{2}}+...+ 
\frac{n!}{q^{n+1}}\right) ,  
\label{Debye}
\end{equation}
which is valid for $x>0$ and $n\geq 1$. For $k>\frac{D-3}{2}$, which
corresponds to $n<1$ in the preceding equation, 
the exponent of $\tau $ in eq.(\ref{Debye}) becomes negative and the integral is
undefined. In this case for small values of $L$, a generalization to
negative odd powers of the argument of the integrand in the Debye integral
can be done using the methods of ref. \cite{svai} and the integral has the
expansion ($u=m_{R}(L)L$ and $n=D-3-2k$), 
\begin{equation}
J(u,n)=\int_{u}^{\infty }d\tau \frac{\tau ^{-n}}{e^{\tau }-1}
=-\sum_{q=0,q\not=n}^{\infty }\frac{B_{q}}{q!}\frac{u^{q-n}}{q-n}-\frac{1}{ 
n! }B_{n}\ln u+\alpha _{n}
\label{J}
\end{equation}
where $B_{k}$ are the Bernoulli numbers and $\alpha _{n}$ is a constant.
Replacing the above equations in Eq.(\ref{mDyson2}) we have in the small $L$
regime the following expression, 
\begin{equation}
m_{R}^{2}(L)=m^{2}+\frac{4\lambda (N+2)}{NL^{D-2}}\left[ A(L,D)+B(L,D)\right]
,  \label{mDyson3}
\end{equation}
where, 
\begin{equation}
A(L,D)=\sum_{k=0}^{k\leq \frac{D-3}{2}
}g(D,k)(m_{R}(L)L)^{2k}I(m_{R}(L)L,D-3-2k)
\end{equation}
and 
\begin{equation}
B(L,D)=\sum_{k>\frac{D-3}{2}}^{k=\infty
}g(D,k)m_{R}(L)L)^{2k}J(m_{R}(L)L,D-3-2k)  \label{B}
\end{equation}
The equations above give a non-perturbative expression for the $L$-corrected
renormalized mass in the small $L$ regime, for even dimensional Euclidean
space. Closed expressions are very difficult to obtain in arbitrary dimension,
but for $L$ sufficiently small, the series in Eq.(\ref{B}) and the
representations (\ref{DEbye}), (\ref{J}) for the functions
$I(m_{R}(L)L,D-3-2k)$ and $J(m_{R}(L)L,D-3-2k)$ could be truncated, giving an
approximate 
algebraic equation for the $L$-corrected mass. The resulting expression should
be injected in eqs.(\ref{G}) and (\ref
{lambda}) to give the $L$-corrected coupling constant in the small $L$
regime.
We will not perform these manipulations
here. Instead, we will in the following obtain some exact results in $D=3$.

In dimension $D=3$, using Eqs.(\ref{Abramov}) the sum over $n$ in Eq.(\ref
{mDyson}) can be performed exactly. We obtain a closed transcendent equation
satisfied by the $L$-corrected renormalized mass, 
\begin{equation}
m_{R}^{2}(L)=m^{2}-\frac{\lambda (N+2)}{N\pi L}\log (1-e^{-m_{R}(L)L}),
\label{mDyson4}
\end{equation}
or, in the large $N$ limit, 
\begin{equation}
m_{R}^{2}(L)=m^{2}-\frac{\lambda }{\pi L}\log (1-e^{-m_{R}(L)L}).
\label{mDyson5}
\end{equation}
The large $N$ renormalized mass is plotted as a function of $L$ in $Fig.3$

Also, as in the preceding section, we have in $D=3$ an exact expression for
the coupling constant in the large $N$ limit as a function of the
renormalized $L$-dependent mass, 
\begin{equation}
\lambda _{R}(D=3,L)=\frac{8\pi m_{R}(L)\lambda (e^{m_{R}(L)L}-1)}{8\pi
m_{R}(L)(e^{m_{R}(L)L}-1)+3\lambda }.  
\label{lambdaexata1}
\end{equation}
In $Fig.4$ we plot in dimension $D=3$ the $L$-corrected coupling constant
for the non-Wick-ordered model using jointly Eqs.(\ref{mDyson5}) and (\ref
{lambdaexata1})
in the same scale we have used to plot the Wick-ordered coupling constant. 
We see that the behaviour of the coupling constant is quite
different from the monotonically increasing in $\frac{1}{L}$ behaviour of the 
renormalized squared
mass and also from the behaviour of the coupling constant in the Wick-ordered
case.
The non-Wick-ordered coupling constant slightly decreases for decreasing
values of $L$ until some minimum value and then starts to increase.
In the Wick-ordered model the
coupling constant tends monotonically to zero as $L$ goes to zero, while in
the non-Wick ordered model it has a non vanishing value even for
very small values of $L$. 
In fact, numerical analysis of the solution of Eq. (\ref{mDyson5}) shows
that $m_{R}(L)L\rightarrow 0$ and $m_{R}^{2}(L)L\rightarrow \infty $ as $ 
L\rightarrow 0$ and, therefore, the $L$-corrected non-Wick-ordered coupling
constant has a non-vanishing value at $L=0$, which is equal to the free
space value $\lambda $. As a general conclusion it can be said that for the
non-Wick-ordered model the $L$-dependent renormalized coupling constant
slightly departs to lower values, from the free space coupling constant.
Furthermore this departure is smaller for smaller values of $\lambda $ (in
units of $m$).

In what concerns the mass behaviour, let us refer to Eq.(\ref{mDyson}) and
remember that the Bessel function $K$ is positive for real positive values
of the argument and decreases for increasing values of the argument. 
For space dimension $D>2$ the correction in $L$ to the
squared mass is positive and the $L$-dependent squared mass is a
monotonically increasing function of $\frac{1}{L}$. If we start in the
disordered phase with a negative squared mass, the model exhibits
spontaneous symmetry breaking of the $O(N)$ symmetry to $O(N-1)$, but for a
sufficiently small critical value of $L$ the symmetry is restored. The
critical value of $L$, $L_{c}$ is defined as the value of $L$ for which the
inverse squared correlation length , $\xi ^{-2}(L,{\bf \varphi }_{0})$,
vanishes in the gap equation, 
\begin{equation}
\xi ^{-2}(L,{\bf \varphi }_{0})=m^{2}+2\lambda {\bf \varphi }_{0}^{2}+\frac{
2\lambda (N+2)}{NL}\sum_{n}\int \frac{d^{D-1}k}{(2\pi )^{D-1}}\;\frac{1}{ 
{\bf k}^{2}+\omega _{n}^{2}+\xi ^{-2}(L,{\bf \varphi }_{0})},  \label{gap}
\end{equation}
where ${\bf \varphi }_{0}$ is the normalized vacuum expectation value of the
field (different from zero in the ordered phase). In the neighbourhood of
the critical point $\varphi _{0}$ vanishes and the gap equation reduces to
Eq.(\ref{mDyson}). If we limit ourselves to the small $L$ regime, we may use
an asymptotic formula for small values of the argument of the Bessel
functions in Eq. (\ref{mDyson}). Then taking $m_{R}(L)=0$ in the resulting
equation it is not difficult to obtain the critical value of $L$ in
Euclidean space dimension $D$ ($D>2$), 
\begin{equation}
(L_{c})^{D-2}=-\frac{4\lambda (N+2)g(D)}{m^{2}N},  \label{critico}
\end{equation}
where 
\begin{equation}
g(D)=\frac{1}{4\pi ^{\frac{D}{2}}}\Gamma (\frac{D}{2}-1)\zeta (D-2),
\end{equation}
$\zeta (D-2)$ being the Riemann $zeta$-function. For $D=3$ the $zeta$
-function in $g(D)$ has a pole and a renormalization procedure implies that
the quantity $L_{c}^{D-2}$ is proportional to the regular part of the
analytic continuation of the $zeta$-function in the neighbourhood of the
pole. It is not difficult to show that in this case the critical value of $L$
is given by the solution of the equation, 
\begin{equation}
L_{c}^{-1}\log (\mu ^{2}L_{c}^{2})=\frac{2\pi m^{2}N}{\lambda (N+2)},
\end{equation}
where the parameter $\mu $ corresponds to a fixed renormalization scale. The
result above generalizes to a phase transition associated to a spacial
boundary, estimates and numerical simulations in \cite{Einhorn} and \cite
{Bimonte}.

\section{Concluding Remarks}

Taking the Wick-ordering, which eliminates all contributions from 
the tadpoles, we decouple in some sense the boundary behaviour 
of the coupling constant from the mass boundary behaviour.  Wick-ordering is
a useful and simplifying
procedure in the applications of field theory to particle physics, but 
the same is not necessarily 
true in applications of field theory to investigate 
critical phenomena, where the contribution from tadpoles could be  
physically significant. As a consequence of the suppression of 
Wick-ordering the boundary behaviour of the coupling constant 
is sensibly modified with respect to the monotonic behaviour in 
the Wick-ordered case (see comments following Eq.(\ref{lambdaexata1})).
As a general qualitative conclusion it can be said that,  
for the 
non-Wick-ordered model in the perturbative regime in $\lambda$,
 the $L$-dependent renormalized coupling constant only slightly
departs from the free space value $\lambda$ and even so, for small values of
$L$.

\section{Acknowledgments}

We are indebted to A.E. Santana for very stimulating and clarifying discussions.
This paper was supported by CNPq (Brazilian National Research Council) and
FAPERJ (Foundation for the support of research in the state of Rio de
Janeiro - Brazil). One of us (A.P.C.M.) is grateful for kind hospitality to
Instituto de Fisica da UFBA (Brazil) where part of this work has been done.

\bigskip \newpage

{\bf Figure captions}

Fig.1 - Typical diagram contributing to the four-point function at leading
order in $\frac{1}{N}$. To each vertex there is a factor $\frac{\lambda }{N}$
and for each single bubble a colour circulation factor $N$

Fig.2 - Renormalized coupling constant (in units of $\frac{1}{N}$) for the
Wick-ordered model as a function of the separation between the planes in
dimension $D=3$.

Fig.3 - Renormalized mass for the non-Wick-ordered model as a function of the 
spacing $L$ between the planes

Fig.4 - Renormalized coupling constant (in units of $\frac{1}{N}$) for the
non-Wick ordered model as a function of the distance between the planes in
dimension $D=3$.


\begin{thebibliography}{99}
\bibitem{Lawrie1}  I.D. Lawrie, Phys. Rev. B {\bf 50}, 9456 (1994)

\bibitem{Lawrie2}  I.D. Lawrie, Phys. Rev. Lett. {\bf 79}, 131 (1997)

\bibitem{Brezin}  E. Br\'{e}zin, D.R. Nelson, A. Thiaville, Phys. Rev. B 
{\bf 31}, 7124 (1985)

\bibitem{NAdolfo1}  A.P.C. Malbouisson, N.F. Svaiter, J. Math. Phys. {\bf 37}
, 4352 (1996)

\bibitem{NAdolfo2}  A.P.C. Malbouisson, N.F. Svaiter, Physica A {\bf 233},
573 (1996)

\bibitem{Gino1}  G.N.J. A\~{n}a\~{n}os, A.P.C. Malbouisson, N.F. Svaiter,
Nucl. Phys. B {\bf 547}, 221 (1999)

\bibitem{Abrikosov}  A.A. Abrikosov, Zh. Eskp. Teor. Fiz. {\bf 32}, 1442
(1957)

\bibitem{Cesar}  C.D. Fosco, A. Lopez, Nucl. Phys. B {\bf 538}, 685 (1999)

\bibitem{FAdolfo}  L. Da Rold, C.D. Fosco, A.P.C. Malbouisson, "Fermionic
determinant with domain wall in 2+1 dimensions", hep-th/0107230 (2001)

\bibitem{Zinn}  J. Zinn-Justin, Quantum Field Theory and Critical Phenomena,
(Clarendon Press, Oxford, 1996)

\bibitem{Elizalde}  A. Elizalde, E. Romeo, J. Math. Phys. {\bf 30}, 1133
(1989)

\bibitem{Abramowitz}  M. Abramowitz, I.A. Stegun, eds., Handbook of
Mathematical Functions, (Dover, New York, 1965)

\bibitem{jac}  L. Dolan and R. Jackiw, Phys. Rev. D {\bf 9}, 3320 (1974)

\bibitem{tet}  M. Reuter, N.Tetradis and C. Wetterich, Nucl. Phys. B {\bf  
401 }, 567 (1993)

\bibitem{svai}  B.F. Svaiter and N.F. Svaiter, J. Math. Phys. {\bf 32}, 175
(1991)

\bibitem{Einhorn}  M.B. Einhorn, D.R.T. Jones, Nucl. Phys. B {\bf 392}, 611
(1993)

\bibitem{Bimonte}  G. Bimonte, D. I\~{n}iguez, A. Taracon, C.L. Ullod, Nucl.
Phys. B {\bf 490}, 701 (1997)
\end{thebibliography}
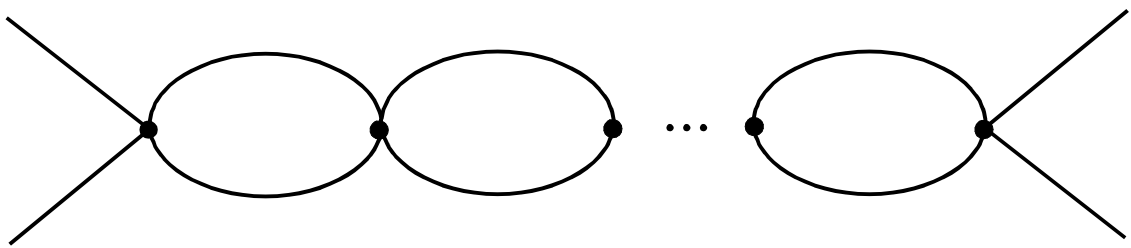
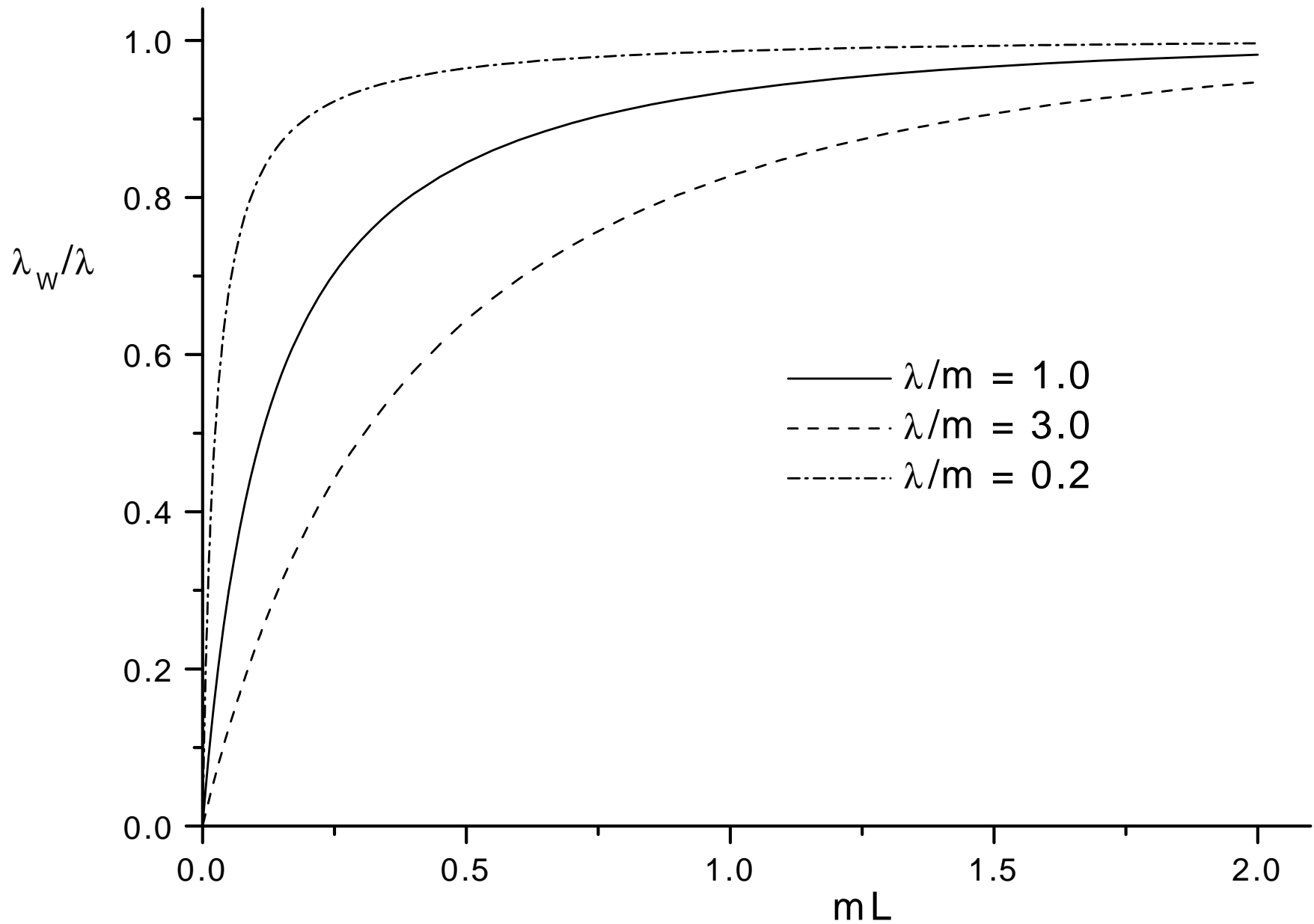
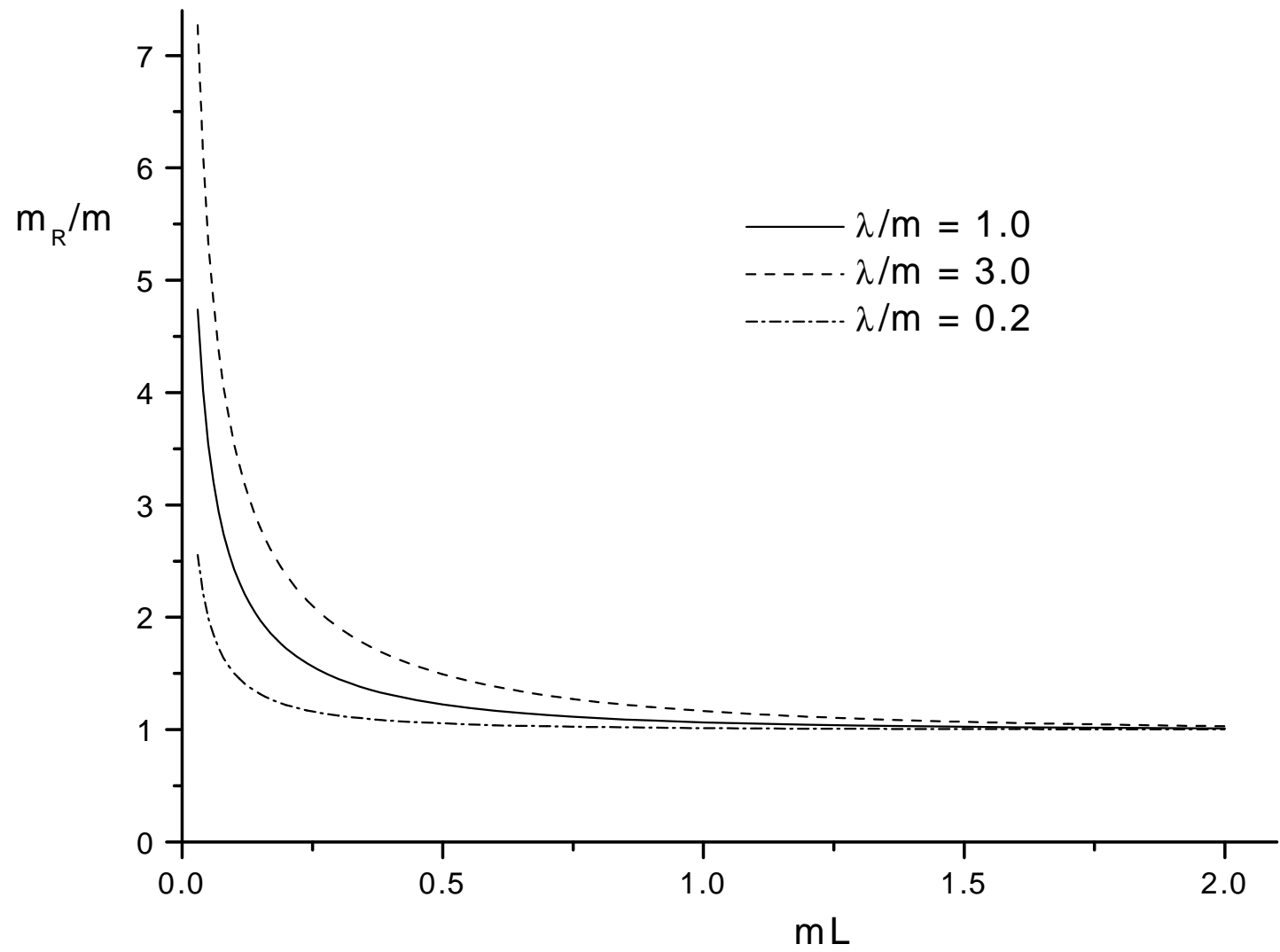
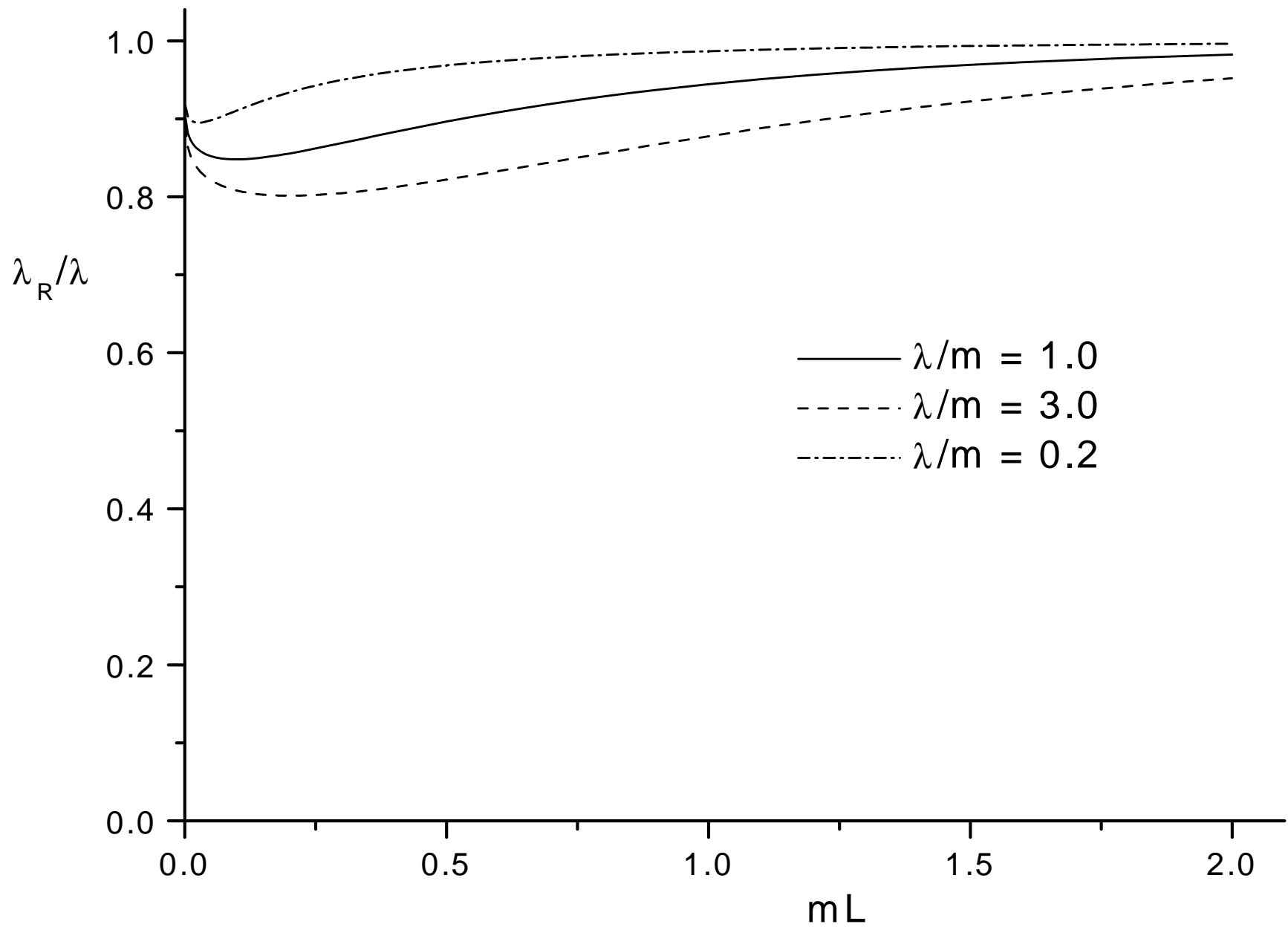
\end{document}